\documentclass[pdflatex,sn-basic]{sn-jnl}% Basic Springer Nature Reference 

\usepackage{graphicx}%

\usepackage{amsmath,amssymb,amsfonts}%
\usepackage{amsthm}%
\usepackage[title]{appendix}%
\usepackage{xcolor}%
\usepackage{textcomp}%
\usepackage{manyfoot}%
\usepackage{booktabs}%
\usepackage{algorithm}%
\usepackage{algorithmicx}%
\usepackage{algpseudocode}%
\usepackage{listings}%
\usepackage[center]{caption}
\DeclareMathOperator*{\argmin}{arg\,min}

\begin{document}

\title[Optimizing substructure search]{Optimizing substructure search: a novel approach for efficient querying in large chemical databases}


\author[1]{\fnm{Vsevolod} \sur{Vaskin}}\email{vsevolod.vaskin@quantori.com}
\equalcont{These authors contributed equally to this work.}

\author[1]{\fnm{Dmitri} \sur{Jakovlev}}\email{dmitriy.yakovlev@quantori.com}
\equalcont{These authors contributed equally to this work.}

\author[1]{\fnm{Fedor} \sur{Bakharev}}\email{fedor.bakharev@quantori.com}
\equalcont{These authors contributed equally to this work.}

\affil[1]{\orgname{Quantori}, \orgaddress{ \city{Cambridge}, \state{Massachusetts}, \country{United States of America}}}

\abstract{Substructure search in chemical compound databases is a fundamental task in cheminformatics with critical implications for fields such as drug discovery, materials science, and toxicology. However, the increasing size and complexity of chemical databases have rendered traditional search algorithms ineffective, exacerbating the need for scalable solutions. We introduce a novel approach to enhance the efficiency of substructure search, moving beyond the traditional full-enumeration methods. Our strategy employs a unique index structure: a tree that segments the molecular data set into clusters based on the presence or absence of certain features. This innovative indexing mechanism is inspired by the binary Ball-Tree concept and demonstrates superior performance over exhaustive search methods, leading to significant acceleration in the initial filtering process. Comparative analysis with the existing Bingo algorithm reveals the efficiency and versatility of our approach. Although the current implementation does not affect the verification stage, it has the potential to reduce false positive rates. Our method offers a promising avenue for future research, meeting the growing demand for fast and accurate substructure search in increasingly large chemical databases.}




\maketitle

\section{Introduction}
Substructure search in chemical compound databases is a vital task in cheminformatics, underpinning broad applications in drug discovery, 
materials science, and toxicology. The objective is to identify all molecules in a database that contain a given query substructure, which
typically corresponds to a specific chemical motif or functional group. This search has been a cornerstone in understanding the influence
of specific substructures on the biological activity, physicochemical properties, and reactivity of a compound, a concept recognized for
decades \cite{barnard1993}.

Historical advancements in computer-assisted substructure search were marked by the adoption 
of graph-based methods to detect particular substructures in chemical compounds. 
A significant milestone was the introduction of the Simplified Molecular Input Line 
Entry System (SMILES) notation by Weininger in the 1980s \cite{weininger1988}. This 
innovation offered a straightforward linear portrayal of molecular configurations in string format.

The search for a structure fundamentally relies on the solution of the subgraph isomorphism problem, a problem known to be NP-complete \cite{ullmann1976}. 
Due to its high computational complexity, numerous algorithms and heuristics have been devised to accelerate the search process. 
Among these, the Filter-and-Verification paradigm has been a prevalent approach, involving an initial filtering step to quickly
eliminate unsuitable candidate graphs, and a more computationally intensive verification step to confirm the presence of the query
substructure in the remaining candidates \cite{cordella2004, shasha2002}. Over time, graph-based subgraph isomorphism algorithms, 
such as the Ullmann algorithm \cite{ullmann1976} and the VF2 algorithm \cite{cordella2004}, have emerged as more efficient and 
scalable solutions for substructure search in large chemical databases.

In addition to these, frequent subgraph mining algorithms like gSpan \cite{yan2002}, FFSM \cite{kuramochi2001}, and Gaston \cite{nijssen2004} 
have proven valuable in identifying frequently occurring substructures in large sets of chemical compounds. These approaches 
are particularly beneficial for applications such as structure-activity relationship (SAR) analysis and molecular classification.

Efficient filtering techniques often involve the use of binary and quantitative features, or fingerprints, to represent molecular 
structures. These fingerprints facilitate the rapid elimination of graphs that do not contain the specific features required by the
query subgraph, thereby speeding up the substructure search process \cite{bonchi2011, klein2011}.

However, as the number of known molecules and the size of chemical databases have grown significantly, traditional approaches,
which often require a full or nearly full enumeration of candidates, have become increasingly challenging to implement efficiently. 
This development underscores the need for more scalable solutions. The complexity is not just computational but also involves
handling increasingly large data sets that cannot be efficiently processed using traditional methods.

Our work introduces a unique approach to mitigating these challenges. While in certain cases the algorithm may resort to exhaustive
enumeration, in most scenarios it employs a more sophisticated strategy, transcending the conventional full enumeration paradigm. 
Instead, we introduce a unique index structure: a tree that segments the molecular dataset into clusters based on the presence orc
absence of features. Inspired by the binary Ball-Tree concept \cite{omohundro1989, clarkson2006}, this structure demonstrates 
superior performance over exhaustive search on average, leading to a significant acceleration in the filtering process.

We provide a comparative analysis of our method based on Bingo \cite{Pavlov2010} with the original Bingo algorithm, highlighting key differences. 

Although Bingo uses advanced filtering effectively, it relies on exhaustive search in a chemical space that continually expands. On the contrary,
our approach departs from exhaustive search and places an existing molecular fingerprint into a tree structure, rather than a
conventional relational database. While our current version does not impact the verification stage, it speeds up the filtering stage.
By introducing this innovative structure, we aim to cater to the growing scale of chemical databases and the escalating demand
for efficient and scalable search solutions. Our approach offers potential for future research and application in the quest
for more efficient and accurate substructure search techniques.

\section{Algorithm description}

\begin{figure}
\centering
    \includegraphics[width=0.7\textwidth]{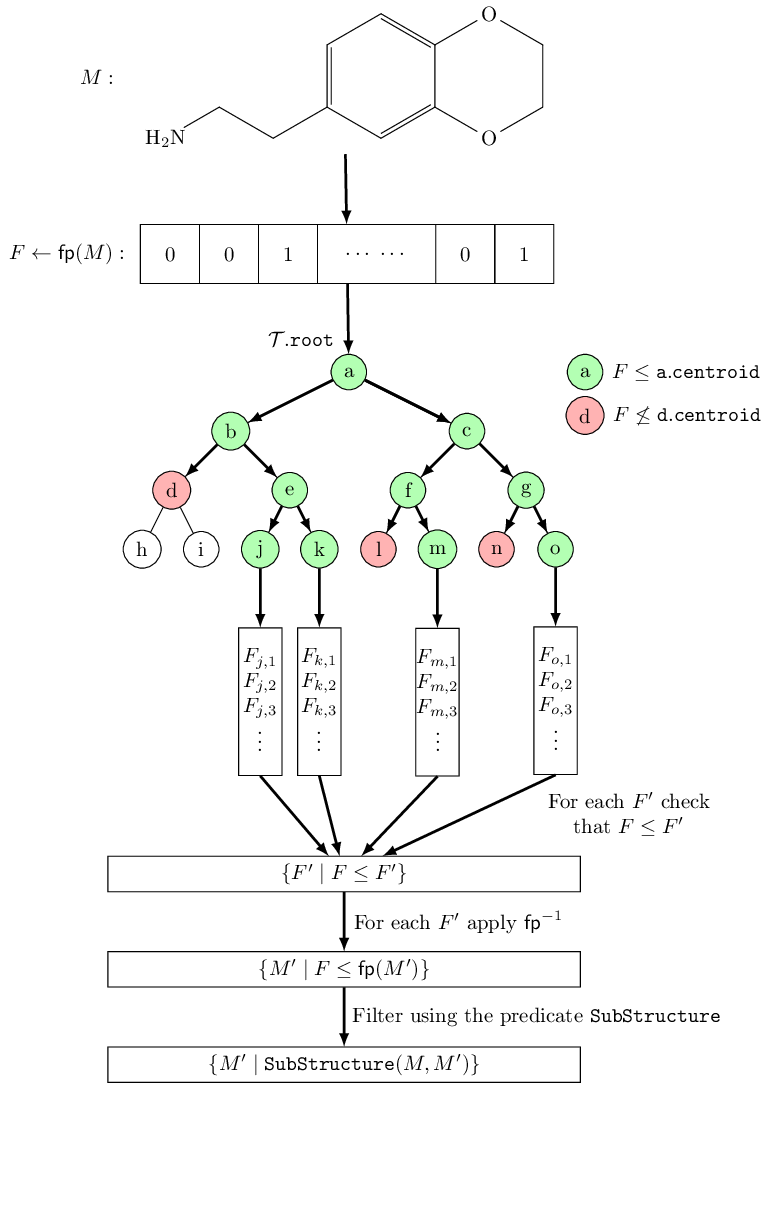}
    \caption{The scheme of the algorithm}
    \label{fig-1}
\end{figure}

\subsection{Notation and main idea}
The objective of our research is to facilitate the identification of specific substructures within molecules from a database $\mathcal{M}$. 
For this, we utilize the concept of a ``fingerprint'', a binary string of constant length ${\sf fl}$, corresponding to each molecule. 
To perform this mapping, we define a function ${\sf fp}: \mathcal{M} \to \mathcal{F}$ that takes a molecule from the set $\mathcal{M}$ and produces its
corresponding fingerprint in the set $\mathcal{F}$. 

To make the substructure search process more efficient, we propose organizing these fingerprints in a binary search tree, denoted $\mathcal{T}$. 
The tree is binary and complete, having a specific depth $d$.

In this tree, the root, left and right subtrees of a node $\tt{v}$ are represented as $\mathcal{T}.\tt{root}$, $\tt{v.left}$, and $\tt{v.right}$, 
respectively. Each node also has a set of all leaves in its subtree, denoted as $\tt{v.leaves}$. Each leaf $\ell$ in the tree $\mathcal{T}$ holds
a set $\ell.\tt{set}$ of fingerprints. A unique concept of our approach is the centroid, ${\tt v.centroid}$, recorded at each node $v$. 
The centroid is defined as a fingerprint $F$ for which $F[i] = 1$ if and only if there exists another fingerprint $F'$ in the subtree
of $\tt{v}$ such that $F'[i] = 1$. This is represented as
$${\tt v.centroid} = \bigvee\limits_{\ell \in {\tt v.leaves}}\bigvee\limits_{F \in \ell.{\tt set}} F.$$ 
This concept of the centroid is inspired by BallTree literature.

Our search process is designed to locate all fingerprints $F'$ in the set $\mathcal{F}$ where $F$ is a submask of $F'$. This search is based on the relation $F_1 \le F_2$ for the fingerprints $F_1, F_2$ that holds if and only if for every $i \in {1, 2, \ldots, {\sf fl}}, \ F_1[i] \le F_2[i]$. 

The search starts from the root and recursively descends into both subtrees. We stop the recursive descent if we reach a node ${\tt v}$ where $F \not\le {\tt v.centroid}$. 
Conversely, if we reach a leaf $\ell$ and $F \le \ell.{\tt centroid}$, we add to $\mathcal{F}_M$ the set $\{F' \in  \ell.{\tt set} \mid {\sf fp}(M) \le F'\}$.

Following the generation of $\mathcal{F}_M$, the next phase involves examining each $$M'\in \bigcup\limits_{F' \in \mathcal{F}_M} {\sf fp}^{-1}(F').$$ The
objective is to determine whether each $M'$ is a substructure of~$M$. This determination is made by employing external algorithms to verify the predicate $\verb|SubStructure|(M', M)$, which is true if and only if $M'$ is a substructure of~$M$. 

Further details on the BallTree and the utilization of the tree in the substructure search process will be provided in the subsequent sections.

The pseudocode for the fingerprint search function in the tree is described in Algorithm \ref{alg:FindInSubtree}. The pseudocode for the function that searches for superstructures of a given molecule is described in Algorithm \ref{alg:FindMetaStructures}.

\begin{algorithm}[ht!]
  \caption{Searching for all matching fingerprints in a subtree}\label{alg:FindInSubtree}
  \begin{algorithmic}[1]
    \Require{${\tt v}$ is a tree vertex, $F$ is a fingerprint}
    \Ensure{$\{F' \in \bigcup\limits_{\ell \in {\tt v.leaves}} \ell.{\tt set} \mid F \le F' \}$}
    \Procedure{FindInSubtree}{${\tt v}, F$} 
    \If{$F \not\le {\tt v.centroid}$} \label{alg:FindInSubtree:line:RecursionCut}
      \State \textbf{return} $\varnothing$
    \ElsIf{${\tt v} \text{ is leaf}$}
      \State \textbf{return} $\{F' \in {\tt v.set} \mid F \le F' \}$ 
    \Else
      \State ${\tt left} \gets $ \Call{FindInSubtree}{${\tt v.left}, F$}  
      \State ${\tt right} \gets $ \Call{FindInSubtree}{${\tt v.right}, F$} 
      \State \textbf{return} \Call{Concatenate}{${\tt left}, {\tt right}$} 
    \EndIf
    \EndProcedure
  \end{algorithmic}
\end{algorithm}

\begin{algorithm}[!ht]
  \caption{Searching for all superstructures of a given molecule} \label{alg:FindMetaStructures}
  \begin{algorithmic}[1]
    \Require $M $ is a molecule 
    \Ensure $\{M' \in \mathcal{M} \mid {\tt SubStructure}(M, M') \}$ 
    \Procedure{FindMetaStructures}{$M $}
    \State $F \gets {\sf fp}(M) $ 
    \State $F_M \gets $ \Call{FindInSubtree}{$\mathcal{T}.{\tt root}, F$}
    \State \textbf{return} $\{M' \in \bigcup\limits_{F' \in \mathcal{F}_M} {\sf fp}^{-1}(F') \mid \Call{\tt SubStructure}{M, M'}\}$ 
    \EndProcedure
  \end{algorithmic}
\end{algorithm}

\subsection{Building the tree}

\begin{algorithm}[ht!]
  \caption{Building the tree} \label{alg:BuildTree}
  \begin{algorithmic}[1]
    \Require $\mathcal{F}$ is the set of all fingerprints, $d$ is the depth of the tree
    \Ensure $\mathcal{T} $ is the BallTree for the superstructure fingerprint search 
    \Procedure{BuildTree}{$\mathcal{F}, d$}
      \State ${\tt v} \gets$ new node
      \If{$d = 1$} 
	\State ${\tt v.set} \gets \mathcal{F}$ 
	\State ${\tt v.centroid} \gets \bigvee\limits_{F \in \mathcal{F}} F$ 
	\State \textbf{return} ${\tt v}$ 
      \Else 
        \State $\mathcal{F}_l, \mathcal{F}_r \gets $ \Call{SplitFingerprints}{$\mathcal{F}$}
        \State ${\tt v.left} \gets $ \Call{BuildTree}{$\mathcal{F}_l, d - 1$} 
	\State ${\tt v.right} \gets $ \Call{BuildTree}{$\mathcal{F}_r, d - 1$}
	\State ${\tt v.centroid} \gets {\tt v.left.centroid} \lor {\tt v.right.centroid}$ 
        \State \textbf{return} ${\tt v}$ 
      \EndIf
    \EndProcedure
  \end{algorithmic}
\end{algorithm}

To start, let us create a trivial tree with a single node, denoted as $\mathcal{T}.{\tt root}$. Assign $\mathcal{T}.{\tt root.set} = \mathcal{F}$. 
Next, we will inductively split the leaves of the tree into two parts, thereby adding new nodes to the tree.

More formally, for each leaf node $\ell$ of the tree, we will divide $\ell.{\tt set}$ using a specific function called
{\tt SplitFingerprints}: $\mathcal{F}_l, \mathcal{F}_r \gets {\tt SplitFingerprints}(\ell.{\tt set})$ ($\mathcal{F}_l \sqcup \mathcal{F}_r = \ell.{\tt set}$).
Next, we will recursively build trees for $\ell.{\tt left}, \ell.{\tt right}$ using the sets $\mathcal{F}_l, \mathcal{F}_r$.

We will continue to split the leaves in this manner until $\mathcal{T}$ becomes a full binary tree with depth $d$. The pseudocode
for the algorithm described above can be found in~\ref{alg:BuildTree}.

\begin{algorithm}
  \caption{Algorithm for splitting fingerprints in parts during tree construction} \label{alg:SplitFingerprints}
  \begin{algorithmic}[1]
    \Require set $\mathcal{F}$ of fingerprints to be split
    \Ensure the split $\mathcal{F}_l, \mathcal{F}_r$ of the set $\mathcal{F}$
    \Procedure{SplitFingerprints}{$\mathcal{F} $}
      \State $j \gets \argmin\limits_{i}\{ \left| |\mathcal{F}| - 2k \right| \mid k = \# \{F \in \mathcal{F} \mid F_i = 1 \} \}$ %\Comment{{\color{red} стоит ли пояснить формулу?}}
      \State $\mathcal{F}_l \gets \{F \in \mathcal{F} \mid F[j] = 0\}$
      \State $\mathcal{F}_r \gets \{F \in \mathcal{F} \mid F[j] = 1\}$ 
      \If {$|\mathcal{F}_l| > \lfloor \frac{n}{2} \rfloor$}
	\State $\mathcal{F}_r \gets \mathcal{F}_r \ \cup$ \Call{TakeLastElements}{$\mathcal{F}_l, |\mathcal{F}_l| - \lfloor \frac{n}{2} \rfloor$}
	\State $\mathcal{F}_l \gets $ \Call{DropLastElements}{$\mathcal{F}_l, |\mathcal{F}_l| - \lfloor \frac{n}{2} \rfloor$} 
      \ElsIf{$|\mathcal{F}_r| > \lceil \frac{n}{2} \rceil$}
	\State $\mathcal{F}_l \gets \mathcal{F}_l \ \cup$ \Call{TakeLastElements}{$\mathcal{F}_r, |\mathcal{F}_r| - \lceil \frac{n}{2} \rceil$}
	\State $\mathcal{F}_r \gets $ \Call{DropLastElements}{$\mathcal{F}_r, |\mathcal{F}_r| - \lceil \frac{n}{2} \rceil $} 
      \EndIf
      \State \textbf{return} $\mathcal{F}_l, \ \mathcal{F}_r$ 
    \EndProcedure
  \end{algorithmic}
\end{algorithm}

We want to perform the splits in such a way that, on average, the search often prunes branches during the traversal. 
That is, the {\bf if} statement in line \ref{alg:FindInSubtree:line:RecursionCut} of the algorithm \ref{alg:FindInSubtree}
should be executed frequently. Let us discuss the function SplitFingerprints in more detail.

Initially, one might consider selecting a specific bit $j$ and assigning all fingerprints $F$ such that $F[j] = 0$ to 
the left subtree, and those with $F[j] = 1$ to the right subtree. In this case, when searching for superstructures of
the fingerprint $F'$, if $F'[j] = 1$, the entire left subtree would be cropped. However, in practice, this approach leads
to significant differences between the left and right parts after a few splits, making it difficult to create a deep and
balanced tree. Unfortunately, a shallow or unbalanced tree does not offer substantial improvements over a full search, 
as it barely eliminates any search branches.

Therefore, we suggest the following method: we will still select the bit as mentioned above, but we will divide the
fingerprints in a way that ensures the sizes of the resulting partitions match. For instance, if the optimal division 
of $n$ fingerprints yields parts with sizes $n_0, n_1 (n_0 < n_1 \ \land \ n_0 + n_1 = n)$, then all values with zero 
will be assigned to the left partition, while the values with one will be distributed to achieve final left and right 
partition sizes of $\lfloor\frac{n}{2}\rfloor, \lceil \frac{n}{2} \rceil$ respectively. If $n_0 > n_1$, we will proceed
symmetrically. The algorithm for the SplitFingerprints function can be found in the pseudocode \ref{alg:SplitFingerprints}.

\section{Benchmarks}

In this study, we have performed a comprehensive benchmarking to assess the performance of our algorithm, which is an extension
of the Bingo fingerprinting system, compared to the established index, namely Bingo \cite{Pavlov2010}. Notably, throughout the 
benchmarking, search results were constrained to the first 10,000 answers to maintain consistency and manageable data size.
Our benchmarking process was performed under the following conditions:

\begin{itemize}
\item OS: Ubuntu 22.04
\item Processor: Intel Xeon E5-2686 v4 (Broadwell)
\item Clock speed: 2.7 GHz
\item RAM: 120 GB
\end{itemize}

The query dataset used for the benchmarking was retrieved from \url{https://hg.sr.ht/~dalke/sqc/browse?rev=tip}, which contains 3488
relevant queries for the substructure search. Ten queries were excluded due to various issues, resulting in a final set of 3478 compounds.

\begin{table}[ht!]
    \begin{tabular}{|c|c|c|}
\hline
\% & \begin{tabular}{@{}c@{}} { Qtr Algorithm,} \\ {single-threaded,} \\ { in-memory}\end{tabular} 
   & \begin{tabular}{@{}c@{}} {Bingo NoSQL,}\\ { single-threaded}\end{tabular} \\
\hline
10\% & 0.0273381 & 0.526846 \\
20\% & 0.053802 & 0.554869 \\
30\% & 0.100528 & 0.610074 \\
40\% & 0.208735 & 0.700541 \\
50\% & 0.553334 & 0.841574 \\
60\% & 0.938981 & 1.06477 \\
70\% & 1.09632 & 1.48609 \\
80\% & 1.36175 & 2.61958 \\
90\% & 2.61875 & 6.42211 \\
95\% & 7.83572 & 13.3279 \\
\hline
$\le$  60 s: & 98.56\% & 98.39\% \\
\hline
\end{tabular}
\caption{\label{bench} Performance Comparison of Single-Threaded In-Memory Execution Between Qtr Algorithm and Bingo NoSQL.}
\end{table}

\begin{figure}
    \centering
\includegraphics[width=0.8\textwidth]{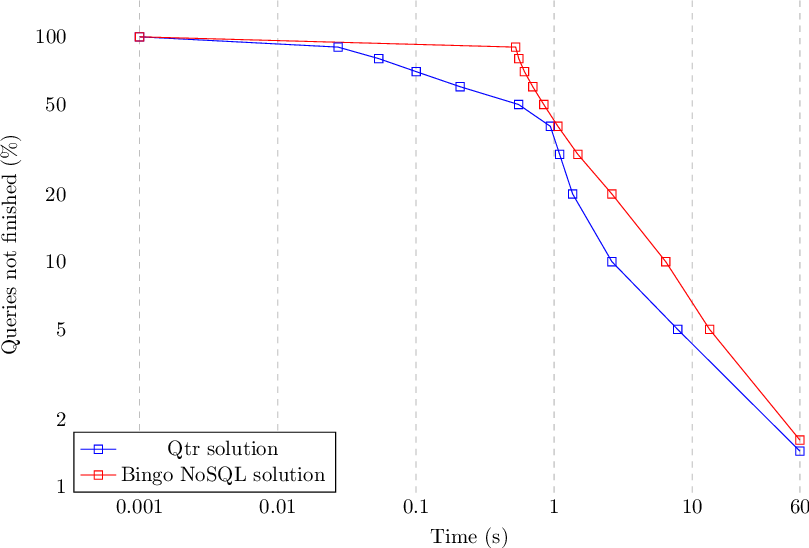}
\caption{Log-Log Plot of Percentage of Unfinished Queries vs Time for Qtr Algorithm and Bingo NoSQL}
\label{fig-2}
\end{figure}

For a single-threaded in-memory execution, our algorithm demonstrates competitive performance. The Table \ref{bench} and Fig. \ref{fig-2} summarizes these benchmark timings, providing a clear comparison between our Qtr algorithm and the Bingo algorithm. This detailed analysis offers valuable information about the performance and potential scalability of our algorithm.

\section{Further Development}

Fingerprints currently form the basis of our algorithm, but they do have certain limitations which do not make them the ideal fit for our tree-based approach.

Firstly, the condensed nature of the fingerprint is aimed at ensuring efficient computation, which often leads to grouping together several characteristics. 
For instance, a single attribute in a fingerprint often encapsulates multiple individual elements because these isolated items, while lacking substantial
filtering power across the entire dataset, might be relevant for specific subsets. However, the fingerprint structure does not account for such instances. 
On the contrary, our approach could accommodate more complex functions, even if they operate slower than traditional filtering methods, for example, using
a fingerprint variant that does not amalgamate different elements.

Secondly, fingerprints are designed to provide a universal filter throughout the dataset. This results in a significantly reduced set of
attributes applicable to the entire database. For example, Bingo utilizes 2584 attributes, which intuitively seem insufficient to capture all the
peculiarities of a 113M-sized molecule dataset. Even a substantially enlarged fingerprint variant would not be able to cover all exceptional cases. 
In contrast, our approach, by dealing with subsets, can extract a unique characteristic for a tree node relevant to the set in the given subtree, 
thus allowing for much more effective coverage of the existing data nuances.

As a result, a potential enhancement of our algorithm might involve the use of a specific attribute in each tree node. Depending on its presence or absence, 
the search continues in both subtrees or only in the right subtree. This attribute would be chosen in advance to approximately bisect the set in the subtree. 
A leaf would contain several characteristics that would be examined when filtering elements from the leaf.

Using the method described above, we could potentially improve the false-positive rate, as the selected attributes would be relevant to the examined subsets. 
Moreover, these attributes could be utilized during verification, possibly resulting in substantial improvements in the verification stage speed, thanks to the
relevance of these attributes to the molecule subsets.

\section{Conclusion}

The current version of our approach can serve as an extension of a fingerprint, enhancing filtering speed by avoiding exhaustive enumeration. Moreover, 
the tree's ability to cluster molecules enables a more detailed examination of cluster-specific attributes, an aspect that existing algorithms struggle with, 
as they aim to find optimal ways to generalize across the entire dataset. Therefore, our approach could potentially be used in the future to improve both the
false-positive rate and the verification speed.

\bibliography{Substructure}

\end{document}